\documentclass[angeo]{copernicus}
\pdfoutput=1

\def\env{\epsilon}

\def\N{\mathbb{N}}
\def\Z{\mathbb{Z}}

\newcommand{\E}[1]{\mathbf{E}\,#1}

\begin{document}

\title{Transmission code optimization method for incoherent scatter radar}

\author[1]{Juha Vierinen}
\author[1]{Markku S. Lehtinen}
\author[1]{Mikko Orisp\"a\"a}
\author[2]{Ilkka I. Virtanen}

\affil[1]{Sodankyl\"a geophysical observatory}
\affil[2]{University of Oulu}

\runningtitle{Transmission code optimization method for incoherent scatter radar}
\runningauthor{Vierinen et.al.}
\correspondence{Juha Vierinen\\ (Juha.Vierinen@iki.fi)}

\received{}
\pubdiscuss{} 
\revised{}
\accepted{}
\published{}
\firstpage{1}

\maketitle

\begin{abstract}
When statistical inversion of a lag profile is used to determine an
incoherent scatter target, the posterior variance of the estimated
target can be used to determine how well a certain set of transmission
codes perform. In this work we present an incoherent scatter radar
transmission code optimization search method suitable for different
modulation types, including binary phase, polyphase and amplitude
modulation. We find that the combination of amplitude and phase
modulation provides better performance than traditional binary phase
coding, in some cases giving better accuracy than alternating codes.
\end{abstract}

\introduction

Incoherent scatter radar lag profile measurements can be deconvolved
using statistical inversion with arbitrary range and time resolution
as shown by \cite{virtanen08}. In this case, the transmission does not
need to be completely free of ambiguities. The only important factor
is the variance of the estimated target autocorrelation function. Even
though alternating codes are transmission sequences that are optimal
in terms of posterior variance when integrated over the code
transmission cycle \citep{lehtinenPhd}, shorter and only slightly less
optimal code groups are beneficial in many cases where an alternating
code sequence is too long. Also, a shorter code group offers much more
flexibility when designing radar experiments, e.g., making it easier
to combine multiple different experiments in the same frequency
channel and simplifying ground clutter removal. The use of short
transmission codes is described in more detail in the companion paper
by \citet{virtanen08a} submitted to this same issue.

We have previously studied the target estimation variance of a
coherent target where the target backscatter is assumed to stay
constant while the transmission travels through the target
\citep{juha08}. We found using an optimization algorithm that a
combination of amplitude and arbitrary phase modulation can achieve
very close to optimum coding (the order $10^{-4}$ less than optimal in
terms of normalized variance). In this study we apply an optimization
method, similar to the one used for coherent targets, to find
transmission codes that minimize the variance of incoherent target
autocorrelation function estimates. We compare results of the
optimization algorithm for several different modulation methods.

All formulas in this paper use discrete time, unless otherwise
stated. All waveforms discussed are complex valued baseband
signals. The ranges will be defined as round-trip time for the sake of
simplicity.

\section{General transmission code}

A code with length $L$ can be described as an infinite length sequence
with a finite number of nonzero bauds with phases and amplitudes
defined by parameters $\phi_k$ and $a_k$. These parameters obtain
values $\phi_k \in [0,2\pi]$ and $a_k \in
[a_{\mathrm{min}},a_{\mathrm{max}}]$, where $k \in [1,\ldots, L] : k
\in \N$. The reason why one might want to restrict the amplitudes to
some range stems from practical constraints in transmission
equipment. Usually, the maximum peak amplitude is restricted in
addition to average duty cycle. Also, many systems only allow a small
number of phases. The commonly used binary phase coding allows only
two phases: $\phi_k \in \{0,\pi\}$.

By first defining $\delta(t)$ with $t \in \Z $ as

\begin{equation}
\delta(t) = \left\{
\begin{array}{llr}
1  & \quad \mbox{when} & \quad t = 0 \\
0  & \quad \mbox{otherwise}, &
\end{array} \right.
\end{equation}

we can describe an arbitrary baseband radar transmission envelope $\env(t)$ as

\begin{equation}
\env(t) = \sum_{k=1}^{L} a_k e^{i\phi_k} \delta(t - k + 1).
\end{equation}

We restrict the total transmission code power to be constant for all
codes of similar length. Without any loss of generality, we set code
power equal to code length

\begin{equation}
L = \sum_{t=1}^{L} |\env(t)|^2.
\end{equation}
This will make it possible to compare estimator variances of codes
with different lengths. It is also possible to compare codes of the
same length and different transmission powers by replacing $L$ with
the relative transmission power.

\section{Lag estimation variance}

We will only discuss estimates of the target autocorrelation function
$\sigma_{\tau}(r)$ with lags $\tau$ that are shorter than the length
of a transmission code (here $r$ is the range in round-trip time, and
it is discretized by the baud length). The lags are assumed to be
non-zero multiples of the baud length of the transmission
code. Autocorrelation function estimation variance is presented more
rigorously in the companion paper by \citet{lehtinen08a} submitted to
the same special issue. The variance presented there also includes
pulse-to-pulse and fractional lags, taking into account target
post-integration as well.

Lag profile inversion is conducted using lagged products for the
measured receiver voltage, defined for lag $\tau$ as 
\begin{equation}
m_{\tau}(t) \equiv m(t)\,\overline{m(t+\tau)}.
\end{equation}

As more than one code is used to perform the measurement, we 
index the codes with $c$ as $\env^c(t)$. For convenience, we 
define a lagged product of the code as
\begin{equation}
\env_{\tau}^c(t) \equiv \env^c(t)\,\overline{\env^c(t+\tau)}.
\end{equation}

With the help of these two definitions, the lagged product measurement
can be stated as a convolution of the lagged product of the
transmission with the target autocorrelation function
\begin{equation}
m^c_{\tau}(t) =  (\env_{\tau}^c*\sigma_{\tau})(t) + \xi_{\tau}(t),
\end{equation}
The equation also contains a noise term $\xi_{\tau}(t)$, which is
rather complicated, as it also includes the unknown target
$\sigma_{\tau}(r)$. This term is discussed in detail, e.g., by
\citet{huuskonen94}. In the case of low SNR, which is typical for
incoherent scatter measurements, the thermal noise dominates and
$\xi_{\tau}(t)$ can be approximated as a Gaussian white noise process
defined as
\begin{equation}
\E{\xi_{\tau}(i)\,\overline{\xi_{\tau}(j)}} = \delta(i-j)\,s^2,
\end{equation}
where $s^2$ is the variance of the measurement noise.

In this case, the normalized measurement ``noise power'' of lag $\tau$
can then be approximated in frequency domain as
\begin{equation}
P_{\tau} \approx \int_0^{2\pi}  \frac{N_c(L - \tau)}{\sum_{c=1}^{N_c}|\hat{\env}_{\tau}^c(\omega)|^2} d\omega,
\end{equation}
where $\hat{\env}_{\tau}^c(\omega) = \mathcal{F}_D^M\{
\env_{\tau}^c(t) \}$ is a zero padded discrete Fourier transform of
the transmission envelope with transform length $M \gg L$. $N_c$ is
the number of codes in the transmission group and $L$ is the number of
bauds in a code. Each code in a group is assumed to be the same length.

will not

For alternating codes of both \citet{lehtinenPhd} and \citet{sulzer93}
type, $P_{\tau} = 1$ for all possible values of $\tau$. For constant
amplitude codes, this is the lower limit. On the other hand, if
amplitude modulation is used, this is not the lower limit anymore,
because in some cases more radar power can be used on certain lags,
even though the average transmission power is the same. 

To give an idea of how phase codes perform in general,
Fig. \ref{optres} shows the mean lag noise power for random code
groups at several different code and code group lengths. It is evident
that when the code group is short and the code length is large, the
average behaviour is not very good. On the other hand, when there is a
sufficient number of codes in a group, the performance is fairly good
even for randomly chosen code groups. Thus, we only need to worry
about performance of short code groups that are sufficiently long.

\section{Code optimization criteria}

Nearly all practical transmission code groups result in such a vast
search space that there is no possibility for an exhaustive
search. As we cannot yet analytically derive the most optimal codes,
except in a few selected situations, we must resort to numerical
means. The problem of finding a transmission code with minimal
estimation variance is an optimization problem and there exist a
number of algorithms for approaching this problem numerically.  

A typical approach is to define an optimization criteria $f(x)$ with a
parameter vector $x$. The optimization algorithm then finds
$x_{\mathrm{min}}$ that minimizes $f(x)$. In the case of transmission
code groups, $x$ will contain the phase $\phi^c_k$ and amplitude
$a^c_k$ parameters of each code in the code group
\begin{equation}
x = \{a_1^1,...,a_L^{N_c},\phi_1^1,...,\phi_L^{N_c}\}.
\end{equation}

There are many different ways to define $f(x)$ in the case of
transmission code groups, but a trivial one is a weighted sum of the
normalized lag power $P_{\tau}$, with weights $w_{\tau}$ selected in
such a way that they reflect the importance of that lag
\begin{equation}
f(x) = \sum_{\tau} w_{\tau}\,P_{\tau}.
\end{equation}

In this paper, we set $w_{\tau} = 1$ for all lags. This gives each lag
an equal importance. This is a somewhat arbitrary choice of weights,
in reality they should be selected in a way the reflects the
importance of the lag in the experiment.

\section{Optimization algorithm}

As our search method will also have to work with codes that have a
finite number of phases, we needed an algorithm that could also work
with situations were an analytic or numerical derivative of $f(x)$
cannot be defined. We developed an algorithm for this specific task,
which belongs to the class of \emph{random local algorithms}
\citep{cstheory97}. Another well known algorithm belonging to the same
class is simulated annealing \citep{simann1}, which has some
similarities to the optimization algorithm that we used.

The random local optimization algorithm is fairly efficient at
converging to a minima of $f(x)$ and it can also to some extent jump
out of local minima. In practice, it is faster to restart the
optimization search with a different random initial parameter set, in
order to efficiently locate the minima of $f(x)$ that can then be
compared the different optimization runs.

A simplified description of our code search algorithm that searches
for local minima of $f(x)$ is as follows:

\begin{enumerate}
\item Randomize parameters in $x$.
\item For a sufficient number of steps, randomize a new value for one
  of the elements of $x$ and accept the change if $f(x)$ is improved.
\item Randomize all parameters $x$, accept the change if $f(x)$ is
improved.
\item If sufficient convergence to a local minima of $f(x)$ has been
  achieved, save $x$ and goto step 1. Otherwise go to step 2. The
  location of the minima can be further fine tuned using
  gradient-based methods, if a gradient is defined for $f(x)$.
\end{enumerate}

In practice, our algorithm also included several tunable variables
that were used in determining the convergence of $f(x)$ to a local
minima. Also, the number of local minima to search for depends a lot
on the number of parameters in the problem. In many cases we are sure
that the global minima was not even found as the number of local
minima was so vast. 

\section{Optimization results}

In order to demonstrate the usefulness of the optimization method, we
searched for code groups that use three different types of modulation:
\emph{binary phase modulation}, \emph{polyphase modulation}, and the
combination of amplitude and polyphase modulation, which we shall
refer to as \emph{general modulation}. In this example, we used $a_k =
1$ for the constant amplitude modulations and allowed amplitudes in
the range $a_k \in [0,2]$ for general modulation codes, while still
constraining the total transmission code power in both cases to be the
same.

\begin{figure*}[t]
\centering
\vspace{-0.1in}
\includegraphics[width=\textwidth]{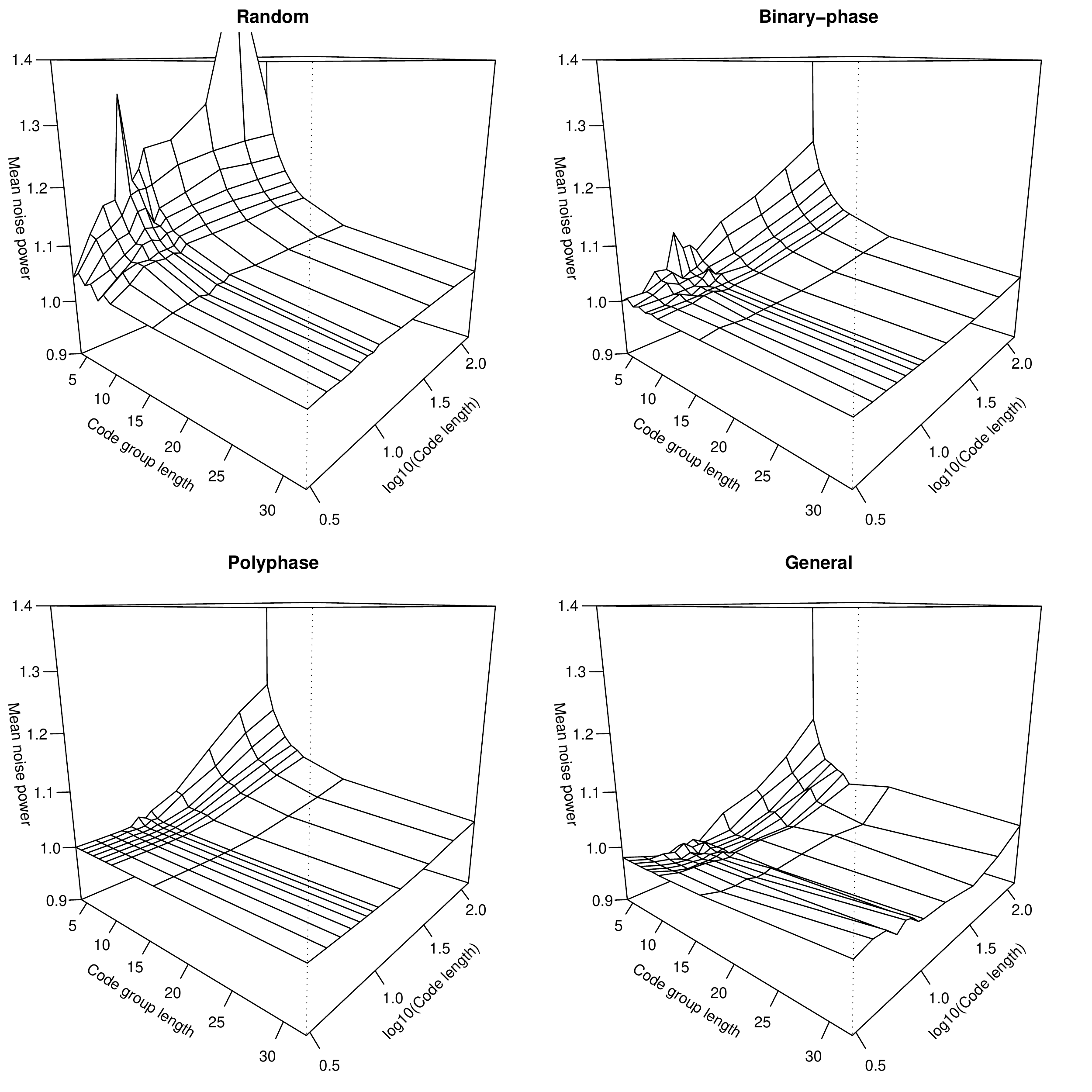}
\caption{The mean lag noise power for random codes, optimized binary
  phase codes, optimized polyphase codes and optimized amplitude $a_k
  \in [0,2]$ and arbitrary phase modulated (general modulation)
  codes. The largest improvements are achieved for short code
  groups. Also, it is clear that the combination of amplitude and
  phase modulation provides the best lag variance.}
\label{optres}
\end{figure*}

The results are shown in Fig. \ref{optres}. In this case the results
are shown in terms of mean lag noise power $P = (L-1)^{-1}\sum_{\tau}
P_{\tau}$. It is evident that significant improvement can be achieved
when the code group length is short. For longer code groups, the
optimized groups do not different that much from random code
groups. Also, one can see that optimized polyphase codes are somewhat
better than binary phase codes; ultimately general phase codes are
better than polyphase codes -- in some cases the mean lag noise power
is actually better than unity. The reason for this is that amplitude
modulation allows the use of more power for measuring some lags, in
addition to allowing more freedom in removing range ambiguities. It
should also be noted, that when the code or code group length is
increased, the difference between modulation methods also becomes less
significant.

\conclusions

We have introduced an optimization method suitable for searching
optimal transmission codes when performing lag profile
inversion. General radar tranmission coding, i.e., modulation that
allows amplitude and arbitrary phase shifts, is shown to perform
better than plain binary phase modulation. Amplitude modulation is
shown to be even more effective than alternating codes, as the
amplitude modulation allows the use of more radar power in a subset of
the lags.

For sake of simplicity, we have only dealt with estimation variances
for lags that are non-zero multiples of the baud length, with the
additional condition that the lags are shorter than the transmission
pulse length. It is fairly easy to extend this same methodology for
more complex situations that, e.g., take into account target
post-integration, fractional or pulse-to-pulse lags. This is done by
modifying the optimization criterion $f(x)$.

In the cases that we investigated, the role of the modulation method
is important when the code length is short. When using longer codes or
code groups, the modulation becomes less important. Also, the need for
optimizing codes becomes smaller when the code group length is
increased. 

Further investigation of the high SNR case would be beneficial and the
derivation of variance in this case would be interesting, albeit maybe
not as relevant in the case of incoherent scatter radar.

\begin{acknowledgements}

This work has been supported by the Academy of Finland (application
number 213476, Finnish Programme for Centres of Excellence in Research
2006-2011). The EISCAT measurements were made with special programme
time granted for Finland. EISCAT is an international assosiation
supported by China (CRIRP), Finland (SA), Germany (DFG), Japan (STEL
and NIPR), Norway (NFR), Sweden (VR) and United Kingdom (PPARC).

\end{acknowledgements}

\bibliographystyle{copernicus}
\bibliography{generalModulationCodes}


\end{document}